\documentstyle[12pt,epsf]{article}

\begin{document}

\begin{flushright}
hep-th/0005241
\end{flushright}

\def\be{\begin{equation}}
\def\ee{\end{equation}}
\def\bea{\begin{eqnarray}}
\def\eea{\end{eqnarray}}

\def\t{\tilde}
\def\cs{(2\pi\alpha')}
\def\CV{{\cal{V}}}
\def\e{{\rm e}}
\def\haf{{\frac{1}{2}}}
\def\tr{{\rm Tr}}
\def\goes{\rightarrow}
\def\goal{\alpha'\rightarrow 0}
\def\gym{g_{{}_{YM}}}
\def\non{\nonumber}

\begin{center}
{\Large{\bf D0-Branes As Confined Quarks \footnote{Talk presented
at {\it Isfahan String Workshop 2000, May 13-14, IRAN.}}}}

\vspace{.5cm}

Amir H. Fatollahi

\vspace{.3 cm}

{\it Institute for Advanced Studies in Basic Sciences (IASBS),}\\
{\it P.O.Box 45195-159, Zanjan, IRAN}

\vspace{.2cm}
{\it and}
\vspace{.2cm}

{\it Institute for Studies in Theoretical Physics and Mathematics (IPM),}\\
{\it P.O.Box 19395-5531, Tehran, IRAN}\\

\vskip .1 in

{\sl fath@theory.ipm.ac.ir}
\end{center}

\begin{abstract}
The possibility of using the quantum mechanics of D0-branes for
the bound-states of quarks and QCD strings is investigated. Issues
such as the inter D0-branes potential, the whiteness of the
D0-branes bound-states and the large-$N$ limit of D0-branes
effective theory are studied. A possible role of the
non-commutativity of relative distances of D0-branes in a study of
ordinary QCD is discussed.
\end{abstract}
D0-branes are defined as particles which strings end on them
\cite{Po1,Po2}. The question is: Can one use \underline{D0-branes
dynamics} as the effective theory of bound-states of quarks and
QCD strings (QCD electric fluxes)? We study the following issues
to approach this question:
\begin{itemize}
\item Inter D0-branes potential to compare with phenomenological one and that of
electric-flux picture.
\item Whiteness of D0-branes bound-states under $SU(N)$ electric
field.
\item Large-$N$ behaviour of D0-branes bound-states to compare with
QCD baryonic states at large-$N$.
\end{itemize}
Also a possible role of involving non-commutativity like the same
one in relative distances of D0-branes in a study of ordinary QCD
is discussed. Discussions here are mostly coming from the results
appeared in \cite{Fat1,Fat2,Fat3}.

Dynamics of $N$ D0-branes is given by the matrix quantum mechanics
resulted from dimensional reduction of $U(N)$ gauge theory to 0+1
dimension, by replacements $A_i\goes X_i$ \cite{W}: \footnote{Here
we ignore supersymmetry. Also we work in arbitrary dimensions
$d$.} \bea\label{1.1} &~&S=\int dt\; m_0 \tr\; \biggl(\haf
D_tX_i^2 +\frac{[X_i,X_j]^2}{4\cs^2}\biggl),\\ &~&i,j=1,...,d,\;\;
D_t*=\partial_t*-i[a_0,*],\nonumber \eea with
$\frac{1}{2\pi\alpha'}$=string tension ($l_s=\sqrt{\alpha'}$ and
$g_s$=string coupling). $X$'s are in the algebra by the usual
expansion $X_i=x_{i(a)} T_{(a)}$, $(a)=1,..., N^2$. \footnote{To
avoid confusion, we put group indices always in ( ).}

The action is invariant under gauge transformations:
\bea\label{gst1}
\vec{X}&\rightarrow&\vec{X'}=U\vec{X}U^\dagger,\nonumber\\
a_0&\rightarrow&a'_0=Ua_0U^\dagger+iU\partial_tU^\dagger, \eea
with $U$ as arbitrary unitary matrix, and consequently one finds:
\bea\label{gst2}
D_t\vec{X}&\rightarrow&D'_t\vec{X'}=U(D_t\vec{X})U^\dagger,\nonumber\\
D_tD_t\vec{X}&\rightarrow&D'_tD'_t\vec{X'}=U(D_tD_t\vec{X})U^\dagger.
\eea D0-branes are presented ``classically" by diagonal matrices
and the action takes the form of $N$ free particles for them:
\bea\label{1.5} S=\int dt \sum_{(a)=1}^N \haf m_0
\dot{\vec{x}}_{(a)}^2. \eea The action is non-relativistic, but
can be used for covariant formulation by Light-Cone Frame (LCF)
interpretation with the following identifications
\cite{BFSS,9704080}: \bea m_0=p^+,\;\; t=x^+,\;\;X_i={\rm
transverse\; directions.} \eea By the scalings \cite{KPSDF} \bea
t\rightarrow g_s^{-1/3} t,\;\; a_0\rightarrow g_s^{1/3} a_0,\;\;
X\rightarrow g_s^{1/3} X, \eea one finds the relevant energy and
size scales as: \bea\label{scales} E \sim g_s^{1/3}/l_s,\;\;\;\;\;
l_{d+2}= g_s^{1/3}l_s. \eea The length $l_{d+2}$ should be
identified  as the fundamental length scale of the covariant $d+2$
dimensional theory which is expected that its LCF formulation is
presented by the action (1). So we take (for $d=2$) $l_{d+2}$ as
the inverse of 4 dimensional QCD mass scale, denoted by
$\Lambda_{QCD}$.

\begin{enumerate}
{\bf \item D0-Branes Potential:} The effective potential between
D0-branes comes from the effect of quantum fluctuations around a
classical configuration, presented here by diagonal matrices. This
work is equivalent with integrating over oscillations of strings
stretched between D0-branes. One-loop effective action is given by
($2\pi\alpha'=1$)\cite{IKKT}: \bea\label{3.5} (\int
§dt)V(X_\mu^{cl}) =\frac{1}{2}\tr\log\bigg(
P_\lambda^2\delta_{\mu\nu}-2iF_{\mu\nu}\bigg)-
\tr\log\bigg(P_\lambda^2\bigg), \eea with \bea
P_\mu*\equiv[X_\mu^{cl},*],\;\;\;\;\;
F_{\mu\nu}*\equiv[f_{\mu\nu},*],\;\;\;\;\;
f_{\mu\nu}\equiv[X_\mu^{cl},X_\nu^{cl}],\non\\ \mu ,\nu
=0,1,...,d,\;\;\;\;\;\;X_0=i \partial _t +a_0,\non\\
P_\lambda^2=-\partial_t^2+\sum_{i=1}^d P_i^2, \;\;\;\;\;\;{\rm
for}\;\;\;\;\; a_0^{cl}=0. \eea
For two static D0-branes at distance $r$ we may take:
\bea\label{3.15} &~&X_1^{cl}=\haf \left( \matrix{ r & 0 \cr 0 & -r
} \right),\;\;\;\; X_0^{cl}=i\partial_t \left( \matrix{ 1 & 0 \cr
0 & 1 } \right),\nonumber\\ &~&a_0^{cl}=X_{i>1}^{cl}=0. \eea So
one finds \bea\label{3.20} P_1=\frac{r}{2}\otimes
\Sigma_3,\;\;\;\;P_0=i\partial_t\otimes 1_4,\;\;\;\; P_{i>1}=0,
\eea with $\Sigma_3*=[\sigma_3,*]$ which has 0, 0, $\pm 2$ as
eigenvalues. Also we find the operator
$P_\lambda^2=-\partial_t^2\otimes 1_4 + \frac{r^2}{4}\otimes
\Sigma_3^2$ as a harmonic oscillator operator with frequency
$\omega\sim r/\alpha'$. One-loop is a good approximation for
$\omega\gg m_0\dot{r}^2$ or $rg_s\gg l_s\dot{r}^2$ which for
$g_s\rightarrow 0$ ($m_0\gg l_s^{-1}$) is satisfied for large
separations and low velocities.

One-loop effective action can be calculated easily to find:
\bea\label{a} V(r) &=&
(\frac{d-1}{2})\tr\log\bigg(P_\lambda^2\bigg) \nonumber\\ &=&-
\;2(\frac{d-1}{2}) \int_0^\infty
\frac{ds}{s}\int_{-\infty}^{\infty} dk_0\;
\e^{-s(k_0^2+r^2)}\nonumber\\ &~&+\; {\rm traces\; independent\;
of\; } r, \eea and after integrations one obtains \bea\label{3.30}
V(r)&=&-\; 2(\frac{d-1}{2}) \int_0^\infty \frac{ds}{s}
(\frac{\pi}{s})^{\haf} \e^{-sr^2} \nonumber\\ &=&\;4 \pi
(\frac{d-1}{2})\; |r|\;\non\\ &~&- \infty\; ({\rm independent
\;of\;} r). \eea which is the linear potential with phenomenology
interests \cite{lucha,rosner}. So the effective theory for the relative
dynamics of two D0-branes is given by: \bea\label{3.36} S=\int dt
\bigg(\frac{1}{2} \frac{m_0}{2} \dot{\vec{r}}^2- 4 \pi
(\frac{d-1}{2}) \frac{|\vec{r}|}{2\pi\alpha'}\bigg) \eea and one
finds the energy scale as $E\sim \alpha'^{-2/3}m_0^{-1/3}\sim
g_s^{1/3}/l_s$, as pointed in eq. (\ref{scales}). By assuming the
dynamics in LCF with the longitudinal momentum $m_0$, we have
$M^2\sim p^+p^-\sim m_0E\sim g_s^{-2/3}l_s^{-2}\sim l_{d+2}^{-2}$,
by eq. (\ref{scales}).

This potential is also true for every pair of D0-branes, and one
can write the effective theory for $N$ D0-branes as: \bea S=\int
dt \bigg(\frac{1}{2} m_0\sum_{(a)=1}^{N} \dot{\vec{r}}_{(a)}^2- 4
\pi (\frac{d-1}{2})\sum_{(a)>(b)=1}^{N}
\frac{|\vec{r}_{(a)}-\vec{r}_{(b)}|}{2\pi\alpha'}\bigg). \eea In a
recent work \cite{krishna} by taking the linear potential in
transverse directions of LCF between the quarks of a baryonic
state, the structure functions are obtained with a good agreement
with observed ones.

One can relate the parameter $1/\alpha'$ in the front of the
potential to gauge theory parameters. To do so one needs a string
theoretic description of the gauge theory in LCF, and the natural
guess for this is ``Light-Cone--lattice gauge theory" (LClgt)
\cite{lclgt}. In LClgt one assumes time direction and one of the
spatial directions to be continuous to define LC variables
$x^{\pm}\sim t\pm z$. Other spatial directions play the role of
transverse directions of LCF which are assumed to be lattices. As
usual in LCF, time is $x^+$ and continuous and so we have a
Hamiltonian formulation \cite{kosu} of lgt \cite{willat}. The
linear potential in LClgt, related to string tension is known to
be: \bea V(r)\sim \frac{\gym^2}{a^2} |\vec{r}|, \eea with $a$ as
the lattice spacing parameter in the transverse directions. Via
this one finds the relation \bea \frac{1}{\alpha'}\sim
\frac{\gym^2}{a^2}, \eea for the parameters.

{\bf \item Whiteness:} To find the charge and colour of D0-branes
bound-states we need to know their dynamics in YM backgrounds. In
the case of electromagnetism there is a simple relation:
\bea\label{lf} m_0\ddot{\vec{x}}=q(\vec{E}_{ext.}+\vec{v}\times
\vec{B}_{ext.}). \eea The concept of gauge invariance here is
understood as the invariance of the equations of motion under the
gauge symmetry transformations. In the case of chromodynamics in
r.h.s. matrices in adjoint representation are placed and so they
transform like: \bea \vec{E}\rightarrow \vec{E}'=U\vec{E}
U^\dagger,\;\;\; \vec{B}\rightarrow \vec{B}'=U\vec{B} U^\dagger.
\eea So we need to replace the l.h.s. with matrices with the
correct behaviour under gauge transformations. Now we have good
candidate for non-commutative coordinates: D0-branes coordinates.
One may write for ``matrix'' coordinates \bea
m_0{\ddot{\vec{X}}}=q(\vec{E}_{ext.}+{\dot{\vec{X}}}\times
\vec{B}_{ext.}), \eea but yet the l.h.s. does not have correct
behaviour under gauge transformations! Here the world-line gauge
symmetry eq. (\ref{gst1}) helps us, to write the generalized
``Lorentz" equation as \bea\label{gle}
m_0D_tD_t\vec{X}=q(\vec{E}_{ext.}+
D_t\vec{X}\times\vec{B}_{ext.}), \eea and now by eq. (3) both sides have
an
equal behaviour under gauge transformations. The space dependence
of the fields is a subtle point, because the coordinates
themselves change under transformation on the gauge fields
$A_{(a)}$'s. Resolving the space dependence may be done by
assuming the D0-branes bound-states very small and then taking the
space dependence of external fields just for the centre-of mass
(c.m.) (see fig. below).
\begin{figure}[h]
\begin{center} \leavevmode \epsfxsize=70mm \epsfysize=50mm
\epsfbox{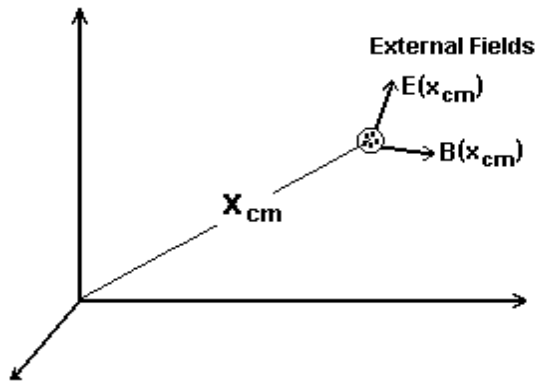} \caption{{\it Space dependence of the
external fields.}}
\end{center}
\end{figure}
The coordinates and momenta of c.m. are given by the trace of
matrices, as: \bea\label{2.55} \vec{x}_{cm}\equiv\frac{1}{N}\;
\tr\vec{X}, \;\;\;\;\;\; \vec{p}_{cm}\equiv \; \tr\vec{P}, \eea
and so are invariant under transformations $X\rightarrow
UXU^\dagger$. So by taking the space dependence just for c.m. we
have: \bea
\vec{E}_{ext}=\vec{E}_{ext}(x_{cm}),\;\;\;\;\;\vec{B}_{ext}=\vec{B}_{ext}(x_{cm}).
\eea

To specify the charge or colour of an extended object (e.g. a
bound-state), we study the dynamics in absence of magnetic field
($\vec{B}=0$) and in uniform electric field
($\vec{E}(x)=\vec{E}_0$). In our case the c.m. dynamics decouples
from non-Abelian parts due to the trace nature of $U(1)$ and
$SU(N)$ parts. So we have: \bea
m_0\ddot{\vec{x}}_{c.m.}=q\vec{E}_{(1)0ext.}, \eea which {\small
(1)} is for $U(1)$ part of $U(N)$. So the dynamics of c.m. will
not be affected by the non-Abelian part: the c.m. is
\underline{white}. It means that each D0-brane sees the net effect
of other D0-branes as the white-complement of its colour: the field
fluxes extracted from one D0-brane to other ones are as the same
of one flux between a colour and an anti-colour.
The linear potential of previous part is consistent with
flux-string picture. The number of D0-branes in the bound-state is
equal to the same of baryons, $N$.

\begin{figure}[h]
\begin{center} \leavevmode \epsfxsize=100mm \epsfysize=50mm
\epsfbox{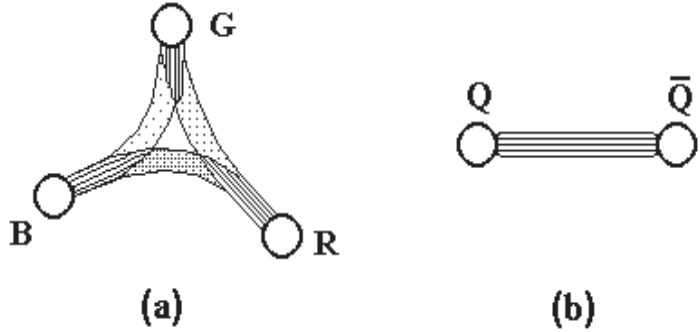} \caption{{\it The net electric flux extracted
from each quark is equivalent in a baryon (a) and a meson (b). The
D0-brane--quark correspondence suggests the string-like shape for
fluxes inside a baryon (a).}}
\end{center}
\end{figure}

{\bf \item Large-$N$:} Baryons show special properties at
large-$N$ limit of gauge theories \cite{witN}:
\begin{itemize}
\item Their mass grows linearly by $N$.
\item Their size is not dependent on $N$. So their density goes to
infinity at large-$N$.
\item Baryon-baryon force grows in proportion to $N$.
\end{itemize}
These properties are mainly extracted from the study of quantum
mechanics of $N$ quarks and their bound-states as an $N$-body
problem. The problem is approached by approximations (e.g.
Hartree) for general potentials which have two characters: 1) they
are attractive, and 2) their strength decreases by 1/$N$ at
large-$N$. Here we check the same behaviours for our problem, by
reminding LCF interpretations. The effective theory for $N$
D0-branes is obtained to be \bea S=\int dt \bigg(\frac{1}{2}
m_0\sum_{(a)=1}^{N} \dot{\vec{r}}_{(a)}^2- 4 \pi
(\frac{d-1}{2})\sum_{(a)>(b)=1}^{N}
\frac{|\vec{r}_{(a)}-\vec{r}_{(b)}|} {2\pi\alpha'}\bigg). \eea
with the relation $1/\alpha'\sim \gym^2/a^2$. Also we have the
replacement at large-$N$: \bea \gym \; \Longrightarrow
\!\!\!\!\!\! \!\!\!\!\!\!
{}^{{}^{large\;N}}\;\frac{\gym}{\sqrt{N}}, \eea and so the action
is read \bea S=\int dt \bigg(\frac{1}{2} m_0\sum_{(a)=1}^{N}
\dot{\vec{r}}_{(a)}^2- 4 \pi
(\frac{d-1}{2})\frac{\gym^2}{a^2}\frac{1}{N} \sum_{(a),(b)=1}^{N}
|\vec{r}_{(a)}-\vec{r}_{(b)}|\bigg). \eea The associated
Hamiltonian of this action is the same used before \cite{witN}
except for the potential term, which is Coulomb one there. Here we
just check the mass: The kinetic term of c.m.
($\frac{\vec{P}^2}{Nm_0}$) grows with $N$, and the net potential
for each D0-brane takes a factor $\frac{1}{2}N(N-1)$ due to pair
interactions. So the potential term grows with
$\frac{1}{2}N(N-1)\gym^2/N\sim N$. The energy grows as $E\sim N$
at large-$N$. In LCF the energy is $P^-$. Also the total
longitudinal momentum of this bound-state is $P^+=Np^+$ with
$p^+=m_0$. So the invariant mass $M$ is read \bea
M^2=2P^+P^--\vec{P}^2\sim N^2\Rightarrow M\sim N. \eea
\end{enumerate}
{\bf Space-Time Considerations: Non-Commutativity}

Relative coordinates of D0-branes are matrices and so
non-commutative. If the correspondence between the dynamics of
D0-branes and confined quarks has a root in Nature, the question
will be about possible justification of this non-commutativity. In
the following 3 comments are in order:
\begin{enumerate}
{\it{\item Special Relativity Idea}}: In the way to find a
consistent theory for the propagation of electromagnetic fields,
special relativity learns to us that space and time should be
treated as a 4-vector $X_\mu$ under boost transformations, such as
the gauge field 4-vector, $A_\mu$.

Also the idea of supersymmetry (SUSY) can be considered as a
natural continuation of the special relativity program: Adding
spin half sector to the coordinates of space-time as the
representative of the fermions of the Nature. This idea leads one
to the super-space formulation of SUSY theories. Also it is the
same way which one introduces fermions to the bosonic string
theory.

Now, what may be modified if in some regions of space and time
there exists non-Abelian (non-commutative) gauge fields? In the
present Nature non-Abelian gauge fields can not make spatially
long coherent states; they are confined or too heavy. But the
picture may be changed inside a hadron. In fact recent
developments of string theories sound this change and it is
understood that non-commutative coordinates and non-Abelian gauge
fields are two sides of one coin. As we discussed, the interaction
between D-branes is the result of path-integrations over
fluctuations of the non-commutative parts of coordinates. It means
that in this picture ``non-commutative'' fluctuations of
space-time are the source of ``non-Abelian'' interactions. One may
summarize this discussion as in the table below: \bea {\rm Field}
& \leftrightarrow & {\rm Space-Time} \non\\ A_\mu\;\;{\rm
(Photons)}&\leftrightarrow & X_\mu \;\;{\rm (4-Vector)\;\;
\;\;\;\;\;\;\;\;\;\;\;\;\;\;\; :Maxwell}\non\\ \psi\;\; {\rm
(Fermions)} & \leftrightarrow & \theta,\;\;\bar\theta\;{\rm
(Super\; Coordinates)\;\;\;\;:SUSY}\non\\ A_\mu^{(a)}\;\;{\rm
(Gluons)}&\leftrightarrow & X_\mu^{(a)}\;\;{\rm (Matrix
\;Coordinates)\;\; \;:QCD}\non \eea

As it has been mentioned previously, the non-commutativity of
D0-branes coordinates just come back to their relative distances
and the c.m.'s of different bound-states of D0-branes presented by
the trace of the position matrices, are commutative objects. We
know that QCD fields are zero outside of hadrons, so the
non-commutativity should be restricted to relative distances
of hadron constituents (see fig.).

\begin{figure}[t]
\begin{center} \leavevmode \epsfxsize=70mm \epsfysize=50mm
\epsfbox{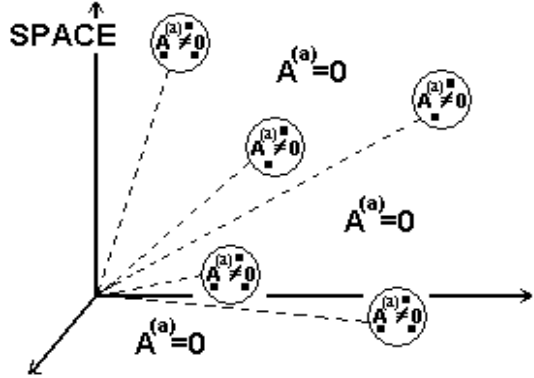} \caption{{\it Non-commutativity inside the
hadrons is just because of the different nature of fields in
them. The c.m.'s are presented by dotted lines.}}
\end{center}
\end{figure}

{\it{\item Recent Example:}} Pure U(1) gauge theory on ordinary
space has free photons. On non-commutative space the theory has
interacting photons and the structure of the theory becomes very
similar to the same of non-Abelian gauge theory, summarized in the
table below \cite{CDS}: \bea {\rm Commutative \; Space}
&\rightarrow& {\rm Non-Commutative \; Space}\non\\  {}[X,X]=0
&\rightarrow& [X,X]=i\theta\non\\ {\rm free\; photons} &
\rightarrow & {\rm interacting\; photons}\non\\ F=\partial A
&\rightarrow& F=\partial A +g\{A,A\}\non\\ F'=F &\rightarrow &
F'=U*F*U^{-1}. \eea The lesson of this example is that one may
cover the aspects of non-Abelian gauge theories by changing the
structure of space-time. It means that by assuming
non-commutativity between the coordinates of space-time one can
get a theory with properties similar to non-Abelian theories.

For our special case the question will be about ``Is the structure
of space-time suggested by D0-brane--quark correspondence
appropriate to cover the non-Abelian structure of $U(N)$ gauge
theories or QCD?"

{\it{\item Lattice Continuum Limit:}} Firstly let us have a look
to the procedure of taking the continuum limit of lattice gauge
theories \footnote{This discussion is borrowed from
\cite{huang}.}. Consider the correlation length $\xi$ between two
plaquettes: \bea \xi=a\;\xi_{latt} \eea which is expressed in
terms of dimensionless parameter $\xi_{latt}$, and lattice spacing
parameter $a$, appearing just as a scale factor. $\xi$ is the
physical quantity and in the continuum limit it should remain
constant, providing:
\bea\xi_{latt}\rightarrow_{\!\!\!\!\!\!\!\!\!\!{}_{{}_{a\rightarrow
0}}} \infty. \eea The correlation function has the behaviour: \bea
G(r)\rightarrow_{\!\!\!\!\!\!\!\!\!\!{}_{{}_{r\rightarrow
\infty}}} e^{-r/\xi}, \eea and by setting $r=na$, the $\xi_{latt}$
does not depends on $a$, becoming a function of coupling constant
$g$: \bea \xi_{latt}=-\lim_{n\rightarrow \infty} \frac{n}{\ln
G(n)}=f(g),\eea which we have for it at a critical value
$g_c$:\bea f(g_c)=\infty. \eea One can find the $g$-dependence of
any physical quantity by the function $f(g)$. Assume $Q$ is a
physical quantity with dimension [length]$^d$; putting
$Q=a^dQ_{latt}$ we have: \bea
Q\xi^{-d}=Q_{latt}\xi_{latt}^{-d}.\eea Assuming $Q\xi^{-d}$ is
finite in the continuum limit we have: \bea
Q_{latt}=C[f(g)]^d,\eea with $C$ as a constant.

One can find the behaviour of the function $f(g)$ at the strong and
weak coupling limit. At strong coupling limit lattice gauge theory
gives the string tension $K$ with dimension [length]$^{-2}$, so:
\bea f(g)=\frac{1}{(a^2K)^2}=\frac{C}{\ln^2 (\kappa g)},\;\;g\gg
1, \eea At weak coupling we have the perturbative result as: \bea
\frac{g^2}{4\pi}=\frac{1}{\gamma_0 \ln (M^2/\Lambda^2)},
\;\;\gamma_0=\frac{33}{12\pi},\eea where $M$ is a mass scale. So
we have: \bea f(g)=C'\e^{[1/(2\gamma_0 g^2)]}=C'
\e^{[8\pi^2/(11g^2)]},\;\;g\ll1. \eea These two behaviours are
plotted in the figure, and one can see that the continuum limit
($f(g)=\infty$) is just gained at $g_c=0$. Based on this, for
every finite value of the coupling constant lattice formulation
does not reach to continuum limit.

On the other hand, we know that the natural framework of
formulation theories on discrete space-time is Non-Commutative
Geometry, with the known examples two-point world or the standard
model of particles on two-sheet world \cite{connes}\cite{hoissen}.
\end{enumerate}

\begin{figure}[h]
\begin{center} \leavevmode \epsfxsize=60mm \epsfysize=50mm
\epsfbox{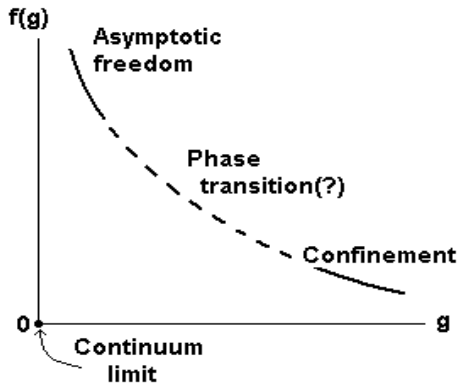} \caption{{\it Plot of $f(g)$ versus coupling
$g$. The continuum limit is just gained at exactly zero coupling.}}
\end{center}
\end{figure}

\end{document}